\documentclass{www2013-accepted}
\usepackage{comment}
\usepackage{amsmath}
\usepackage{caption}
\usepackage{subcaption}
\usepackage{cite}
\usepackage{booktabs}
\DeclareCaptionType{copyrightbox}
\newcommand{\nofr}{\textit{OverallPop}}
\newcommand{\cfr}{\textit{FriendPop}}
\newcommand{\rfr}{\textit{RandFriend}}
\newcommand{\gfr}{\textit{GoodFriend}}
\newcommand{\gcfr}{\textit{GoodFrCount}}

\begin{document}
%
% --- Author Metadata here ---
\conferenceinfo{WWW}{'13}

\title{Do Social Explanations Work? Studying and Modeling the Effects of Social Explanations in Recommender Systems}

\numberofauthors{2} 
\author{
\alignauthor
Amit Sharma\\
       \affaddr{Dept. of Computer Science}\\
       \affaddr{Cornell University}\\
       \affaddr{Ithaca, NY 14853 USA}\\
       \email{asharma@cs.cornell.edu}
% 2nd. author
\alignauthor
Dan Cosley\\
       \affaddr{Information Science}\\
       \affaddr{Cornell University}\\
       \affaddr{Ithaca, NY 14853 USA}\\
       \email{danco@cs.cornell.edu}
}

\date{26 Nov 2012}

\maketitle
\begin{abstract}
Recommender systems associated with social networks often use social explanations (e.g. ``X, Y and 2 friends like this'') to support the recommendations.  We present a study of the effects of these social explanations in a music recommendation context. We start with an experiment with 237 users, in which we show explanations with varying levels of social information and analyze their effect on users' decisions.  We distinguish between two key decisions: the likelihood of checking out the recommended artist, and the actual rating of the artist based on listening to several songs.  We find that while the explanations do have some influence on the likelihood, there is little correlation between the likelihood and actual (listening) rating for the same artist. Based on these insights, we present a generative probabilistic model that explains the interplay between explanations and background information on music preferences, and how that leads to a final likelihood rating for an artist.  Acknowledging the impact of explanations, we discuss a general recommendation framework that models external informational elements in the recommendation interface, in addition to inherent preferences of users.
\end{abstract}



\section{Introduction}
\begin{figure}
\centering
\includegraphics[scale=0.5]{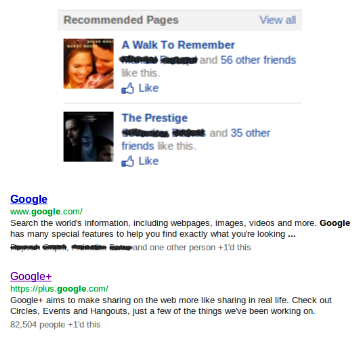}
\caption{Examples of social explanations typically found on the web. Here we show screenshots from Facebook's page recommender and Google. Names and counts of friends, and number of people who like an item are presented to the user.}
\label{fig-google_fb_expl}
\end{figure}

Theories of social proof and social influence \cite{cialdini01} suggest that our preferences are impacted by the actions of those around us. For example, if our friends like a restaurant, we may be tempted to try it out. Many online social networks leverage this notion by supplementing items they recommend with social information about other people who like the item. For example, ``N of your friends like this item'', or ``X and Y recommend this''. 
Figure~\ref{fig-google_fb_expl} shows how these bits of information have permeated our online experience.

Depending on the goals of the system, social information may shed light on the underlying algorithm \cite{symeonidis08} or make recommendations more personal and attractive \cite{herlocker00}. This extra social information can be thought of as an \textit{explanation} for a recommendation.  These explanations may influence how we think about a recommended item. For instance, social explanations might influence people's willingness to try out an item because a trusted friend has endorsed it or they want to be able to talk about it with their friends.  They might influence people's ratings, just as displaying predicted ratings in a recommender system affects people's actual ratings \cite{cosley03}.  They might even influence our opinion of the system itself, by making its decision-making more transparent \cite{sinha02}.

Although these social explanations are becoming popular, little is known about how they affect decision-making around the recommendations, or how users make sense of this social information.  How might different explanations influence our evaluation of a recommended item, and how might particular individuals be more or less susceptible to these explanations?  Further, these explanations often involve disclosing others' interests or past activities and may imply endorsement of the recommended items, raising questions about the acceptability of such explanations.

In the present work, we develop a framework for understanding the effect of social explanations on how people make decisions around recommendations. We first distinguish between two phases of evaluation: before and after consuming a recommended item. In the first phase, a user evaluates her \textit{likelihood} of checking out an item.  In the second phase, the user evaluates the item itself, based on her \textit{consumption} experience.  We can consider these likelihood and consumption ratings as measures of the persuasiveness and informativeness of an explanation: persuasive explanations might increase likelihood ratings, while informative explanations might lead to likelihood ratings that closely align with consumption ratings.  

Through a user study in which we recommend musical artists with social explanations and minimal artist information, we find that different kinds of social explanations do have different effects on likelihood ratings.  However, it is only a secondary effect, with the dominant influence on most people's likelihood ratings being their inherent expectations of how they will like the item.  Further, social explanations are not always persuasive.  Users' comments show that a trusted friend's name can increase the credibility of a recommendation, but a friend whose interests are unknown or incompatible negatively influences likelihood ratings.  Based on these insights, we present a generative model that explains much of the interplay between social explanations and inherent preferences on likelihood ratings, a model that can be generalized to include other sources of explanation as well.  People's comments also revealed that they have quite different strategies for making sense of social explanations, suggesting that personalizing explanations might have real value.

In a second phase of the study, we ask people to return about a week later and listen to music by artists they had rated in the first phase.  We find that the effect of different kinds of social explanations does not transfer to the consumption phase. In fact, like Bilgic et al., we find a low correlation between likelihood and consumption ratings people give to the same artist \cite{bilgic05}.  This suggests that there are different motivations and goals for the two phases, and further, that although explanations are persuasive, they are not very informative and may lead people astray.  

These notions of likelihood and consumption have natural parallels to the ideas of click-throughs and purchases in e-commerce.  The gap between likelihood and consumption suggests that rather than optimizing one or the other, as most recommender work does, it would be fruitful to model both.  We discuss how knowledge of the two phases and their relative characteristics can be used to design recommendation models that consider both the probability of click-throughs and of consumption preferences, helping designers optimize aspects of users' experience to support goals such as serendipity and novelty.

\section{Related Work}

We build on existing work that shows the value of explaining recommendations in general and the growing trend to use social information in recommender systems for both preference modeling and explanation.

\subsection{Explanations in recommender systems} 
Deciding whether to consume a recommended item is not done in isolation, but in a situated context \cite{lueg97}. Terming rating as a cognitive process, Lueg argues that the ratings are a dynamic result of the interaction of an individual with an ``information situation''.  In our context, an explanation is part of the information presented about a recommendation, and studies show that explanations play an important role in helping a user evaluate a recommendation \cite{swear01, tintarev11}. In one of the first studies of explanations, Herlocker et al. evaluated 21 types of explanation interfaces for a movie recommender system \cite{herlocker00}. They found that a histogram showing the ratings of similar users is the most persuasive for users when asked about their likelihood to see a movie.

However, being persuasive has drawbacks.  Another study found that although explanations might persuade a user to try an item, they were not good for accurately estimating the quality of an item \cite{bilgic05}. 
The authors further argue the goal of a recommender should not be to promote a recommendation (which they call \textit{promotion}), but rather enable a user to make a more accurate judgment on the true quality of the item for that person (which they call \textit{satisfaction}).

Besides helping users make an informed choice, explanations may also increase the acceptability of a recommender system overall, by communicating why an item has been recommended to a user \cite{symeonidis08} and thus helping them understand the system.  These explanations and other presentational choices can be designed to increase the system's trustworthiness \cite{pu06}, and a number of real systems incorporate explanations (e.g., Amazon's explanation of ``Customers who bought this also bought these'', and Netflix's explanation by genres).  Tintarev et al. provide a number of desirable attributes of explanations, including transparency, scrutability, trustworthiness, effectiveness, persuasiveness, efficiency, and satisfaction \cite{tintarev07}. 

One outstanding problem it that is not clear how to characterize explanations'  influence on either likelihood or consumption ratings.  Computing persuasiveness is difficult because people's likelihood decisions are also informed by the merits of the recommended item and by other information presented in the interface.  And, though Cosley et al. found that displaying predicted ratings caused people to change their own ratings of movies \cite{cosley03}, this was likely a short-term effect caused by displaying the predicted rating at the same time as the user the rated movie.  Here, we attempt to tease out persuasiveness through comparing a number of different social explanation strategies and by putting a substantial delay between the likelihood and consumption ratings.

\subsection{Social information for recommendation}

With the growth of the social web, systems can use the connections people articulate with people in real life as a source of information for both preference modeling and supporting explanation.  People prefer the use of known friends to explain recommendations over the use of ``similar'' neighbors as computed by many recommendation algorithms \cite{bonhard06}.  This makes sense in the light of literature about how recommendation is a socially embedded process that depends on the relationship and trust between individuals offering and receiving recommendations \cite{lueg97, perugini04}.

Models based on these theories and the availability of social connection information have been proposed to support collaborating filtering algorithms that use social information \cite{konstas09,ma11}, focusing on preferences in users' immediate social networks \cite{guy09, sharma11} and computing trust between people in networks \cite{golbeck06} to improve recommendations.

This information can also be used to support social explanation, as with the neighbor-based ratings in Bilgic and Mooney \cite{bilgic05} and aggregate customer behavior in Amazon. Using user-generated tags, based on their popularity and relevance, is another source of social information that has also been studied for explanation \cite{vig09}.  However, despite the appearance in practice of the use of friendship, egocentric networks, and overall popularity information in social explanations, there has been little study of how they influence likelihood and consumption decisions.  Our work directly addresses these questions, and we now turn to the particular social explanations we study. 

\section{Social Explanations in the Wild} 
\label{sec-social-wild}
Facebook is probably the most ubiquitous context in which we see these social explanations, powered by the Like button. Any page on Facebook, or any entity recognized by Facebook's open graph (such as a movie, a URL, or Facebook content such as status updates or comments), can be Liked. Items in Facebook then present information about who else has Liked them.  A number of other websites dedicated to social recommendation and discovery (such as Getglue or Hunch\footnote{www.getglue.com, www.hunch.com}) suggest items along with an explanation of who (or how many) watched or rated those items. Even in search, such explanations are starting to be shown (Figure~\ref{fig-google_fb_expl}).    

In general, these social explanations follow a few basic forms that theories of social influence suggest might influence people's decision-making \cite{cialdini01}.  In Facebook, for instance, many items have information about how many people in general, or how many of a person's own friends, have Liked the item.  Such explanations rest on the idea of social proof, that people follow other people's behaviors because they assume that others have reasons for doing those things \cite{cialdini98}.  Other social explanations provide the names of particular friends who have Liked the item; particularly if the names chosen are good friends, this might tap into the idea that people we like are more persuasive \cite{christakis09,guy09}.  Finally, the work on trust in recommender systems suggests that recommendations from domain experts are more likely to be persuasive.  Social explanations can also combine these kinds of information, for instance, providing both names and counts of others' activity around items.  

A fundamental question is whether, and how, these social explanations influence user decisions. In addition, we would like to investigate how different types of social information vary in their impact. We are interested in both the \textit{persuasive} power of such explanations, as well as their \textit{informative} power (whether they lead to satisfying choices). From a recommender systems perspective, this leads to questions of how to choose an appropriate explanation for a recommendation, as well as how to choose the recommendations themselves, given desired goals such as end-user satisfaction.  Based on the discussion above, we articulate four high-level research questions:

\noindent \textbf{[RQ1]:} How do different social explanation strategies influence likelihood ratings?

\noindent \textbf{[RQ2]:} How do explanations interact with a person's inherent biases or preferences? 

\noindent \textbf{[RQ3]:} How can we model the effect of explanations on likelihood ratings? 

\noindent \textbf{[RQ4]:} How effective are these explanations in directing people to items that receive high consumption ratings?   

\begin{figure}
        \centering
        \begin{subfigure}[b]{0.19\textwidth}
                \centering
                \includegraphics[scale=0.45]{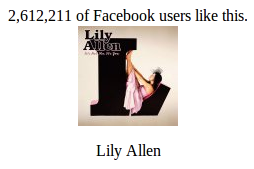}
                \caption{Overall Popularity}
                \label{fig-nofr}
        \end{subfigure}%
        ~
        \begin{subfigure}[b]{0.31\textwidth}
                \centering
                \includegraphics[scale=0.45]{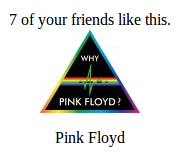}
                \caption{Friend Popularity}
                \label{fig-cfr}
        \end{subfigure}
        
        \begin{subfigure}[b]{0.19\textwidth}
                \centering
                \includegraphics[scale=0.45]{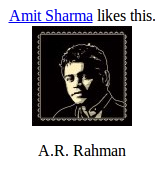}
                \caption{Good/Random Friend}
                \label{fig-grfr}
        \end{subfigure}%
        ~ 
        \begin{subfigure}[b]{0.31\textwidth}
                \centering
                \includegraphics[scale=0.45]{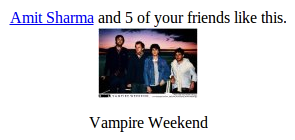}
                \caption{Good Friend \& Count}
                \label{fig-gcfr}
        \end{subfigure}
        \caption{Different explanation strategies used in the experiment, shown along with an artist's name and profile picture. This setup was chosen as a tradeoff between realistic recommendation scenarios (artist information shown) and ideal experiment conditions (no other information).}
        \label{fig-expl_types}
\end{figure}

\begin{comment}
\begin{figure*}
        \centering
        \begin{subfigure}[b]{0.5\textwidth}
                \centering
                \includegraphics[scale=0.3]{graphs/interface1}
                \caption{Phase I: Likelihood}
                \label{fig-phase1}
        \end{subfigure}%
        ~ %add desired spacing between images, e. g. ~, \quad, \qquad etc. 
        \begin{subfigure}[b]{0.5\textwidth}
                \centering
                \includegraphics[scale=0.3]{graphs/interface2}
                \caption{Phase II: Consumption}
                \label{fig-phase2}
        \end{subfigure}
        \caption{Screenshots of ExploreMusic's interface. Explanations are shown in the first phase. The second phase (3 days later) makes users listen to 3 top songs by each artist, but explanations are not shown. The songs were sourced from Grooveshark. }
        \label{fig-interface}
\end{figure*}
\end{comment}

\section{ExploreMusic: A User Study}
We now turn to how we explored these questions through an experiment with 237 users of ``ExploreMusic'', a web application we created that uses Facebook data to explain a series of music recommendations.  We chose music because it is relatively easy to acquire consumption ratings of previously unknown artists (three minutes per song, versus two hours per movie), allowing us to explore whether explanations would influence consumption ratings.  We chose Facebook because it has both social network and music preference information already available: Facebook users Like pages associated with musical artists, which both affirms their preferences and makes them publicly visible by default.

\subsection{Experiment Design}
The experiment took place in two main phases.  We initially collect the artists that the participant and her friends Liked.  We then show all the artists the participant's friends Like that she hasn't yet Liked and ask her to identify a minimum of 30 that she is not familiar with.  We ask for this information  to minimize the effects of prior knowledge.  To minimize position bias, we ordered artists randomly.

\paragraph{Phase I} Phase I begins immediately after the initial selection. The experiment is a within-subjects design, where each participant sees the artists they selected, randomly assigned to one of five explanation strategies: 

\begin{itemize}
\item  \textbf{Overall Popularity}: The number of Likes by all Facebook users for an artist (\nofr, Figure~\ref{fig-nofr}). 
\item \textbf{Friend Popularity}: The number of friends of a user who Like an artist (\cfr, Figure~\ref{fig-cfr}).
\item \textbf{Random Friend}: The name of a particular friend, chosen from those that Like an artist (\rfr, Figure~\ref{fig-grfr}).
\item \textbf{Good Friend}: The name of a ``close'' friend, chosen from those that Like an artist (\gfr, Figure~\ref{fig-grfr}).
\item \textbf{Good Friend \& Count} A combination of Good Friend and Friend Popularity (\gcfr, Figure~\ref{fig-gcfr}).
\end{itemize}

These roughly align with the social explanation strategies described earlier.  
Given a user and an item, \nofr\ and \cfr\ explanations are straightforward to compute using the total number of Facebook users or friends who Like an artist, respectively. For \rfr, we choose a friend at random among all the friends that Like an artist. For \gfr\ and \gcfr, we choose the friend with the highest tie strength who Likes the artist, assuming there exists such a friend with non-zero tie-strength.  Using a rough proxy of interaction frequency, loosely inspired by Gilbert and Karahalios' work on predicting tie strength in Facebook\cite{gilbert09}, we define tie strength between a user and a given friend as the number of interactions (likes, comments, and wall posts) between them among the last 500 interactions involving the user.

For each artist, we show the artist's name, their profile picture on Facebook, and the associated explanation.  For \gfr\ and \gcfr, it was often the case that there were no friends with non-zero tie strength who had Liked the item.  In these cases, we skipped the item, leading us to show fewer artists in these conditions; we saw this as preferable to assigning artists that random friends had Liked because we were afraid that might dilute the effects of close friendship.  
For each recommendation, we ask the user how likely is she to check out the recommended artist and how sure is she about her answer. We use a 0-10 (inclusive) Likert scale to collect these answers\footnote{The initial slider value is 5 and participants usually moved the slider, leading to a relative lack of 5 ratings.}.
To reduce order effects of either artist or explanation strategy, we randomize the order of presentation for artists.

Once all artists are shown, the user fills out a questionnaire that asks about their reaction to the explanations: which ones were more convincing or effective and why, and how she used the information presented to think about the recommended items.  We also asked users about how they felt about our using their friends' social information to explain the recommendations, to see if social explanations raised privacy, identity management, or other issues. 

\paragraph{Phase II}   
In the second phase, users listen to songs by a randomly chosen subset of the artists they had rated in Phase I. Explanations are not shown in this phase. We use Grooveshark\footnote{www.grooveshark.com}, a popular music service, to provide the top three songs for a musician, assuming that a musician's best songs are a reasonable representation of the artist.  Since each listening task takes 6-9 minutes, we randomly chose two artists from each explanation strategy from Phase I to keep the experiment between 60 and 90 minutes.  After listening to the songs, we ask the user to rate how much they liked the artist and their surety about the rating.  As before, feedback was collected on a 0-10 Likert scale.

We required participants to wait at least three days between Phase I and Phase II.  The goal of this delay, and of not re-showing the social explanation during Phase II, was to see whether there was a lasting effect of the explanation on people's consumption ratings \cite{cosley03}. Participants could choose their date for Phase II, with an average delay was 5.2 days.

\subsection{Participants and descriptive overview}
Participants were drawn from two on-campus experimental subject pools covering undergraduate and graduate students as well as staff at the university. Participants were compensated with either money or with experiment participation credits required by some courses.
%% DC 12: How much money?
A total of 237 users took part. Out of these, 175 people completed both phases, while the rest completed only Phase I. The gender ratio was $68\%$ female, $32\%$ male and the average age 20.5 years. We collected a total of 4458 ratings for Phase I and 835 for Phase II.

%%Am23: Taking out this table for space.
\begin{comment}
\begin{table}
\centering
\begin{tabular}{|c|l|}
\toprule
Attribute&Value\\ 
\midrule
Total Participants & 237\\ 
Gender & 161 female, 76 male \\  
Avg. Age & 20.5\\ 
Avg. Friends per user& 735.2\\ 
Total Users (participants+friends) & 151,142 \\ 
Total Likes & ~2.4 million \\ 
Phase I Likelihood ratings & 4458 \\ 
Phase II Consumption ratings & 835 \\
\bottomrule
\end{tabular}
\caption{An overview of participants and data collected.}
\label{tab-desc_stats}
\end{table}
\end{comment}

\begin{table}
\centering
\begin{tabular}{lrcc} 
\toprule
\textbf{Explanation Strategy} & \textbf{N} & \textbf{Mean} & \textbf{Std. Dev.}\\ 
\midrule
\cfr & 1203 & 2.12 & 2.42\\ 
\rfr & 1225 & 2.08 & 2.49\\
\nofr & 1191 & 2.36 & 2.69\\
\gfr & 434 & 2.52 & 2.69\\ 
\gcfr & 405 & 2.71 & 2.90\\
\bottomrule
\end{tabular}
\caption{Likelihood ratings for different explanation strategies. Strategies based on good friends have higher ratings.}
\label{tab-likely_rat}
\end{table}

\section{Phase I: Likelihood}
\subsection{Are different social explanations more persuasive on average?}
In this section, we address \textbf{[RQ1]:} How do different social explanation strategies influence likelihood ratings? Table~\ref{tab-likely_rat} shows the mean likelihood ratings for different explanation strategies\footnote{As a reminder, the good friends-based strategies have fewer ratings because many of the items that were randomly assigned to them hadn't been Liked by a good friend and so were skipped.}.  \gcfr\ and \gfr\ have relatively high mean ratings, while \cfr\ and \rfr\ have relatively low ones, suggesting that good friends are more persuasive than counts or random friends. An ANOVA with repeated measures shows that there is a significant difference between the different explanation strategies ($F(4,763) = 4.96, p=0.0006$). A post-hoc Tukey test shows that \gcfr\ is significantly higher than \rfr\ ($p=0.002$) and \cfr\ ($p=0.006$).

Users' qualitative responses give confirmation, explanation, and depth to these differences, showing the importance of good friends and, no matter which explanation strategy, the importance of identifying with the source of the recommendation.  Table \ref{tab-survey2} shows how useful people saw the different information available to them in explanations, based on coding their responses to a question about what aspects of explanations they found most powerful. 

\paragraph{Showing the right friends matters}
The most important source of information was the name of the friend who liked the item: \textit{``The best recommendation was the showing which one of my friends liked a song. I didn't really care when I was vaguely told `2 friends'. It was important to see names because I know some of my friends' music tastes.''} [P78]

Good friends were seen as more influential and informative than others: \textit{``I would only be interested in the recommendations based on people who are relatively close to me (compared to random individuals/acquaintances on my friends list).''} [P23]

This is likely because people are better able to think about whether they know and trust good friends' tastes, as suggested by \cite{perugini04}: \textit{``I found it most powerful when I could see what friend likes the artist. I know what kind of music my friends listen to and that helps me know if I would like the artist or not.''} [P105]

As Table~\ref{tab-survey2} shows, people also trusted those friends more who were perceived to have similar interests, or a good taste in music: \textit{``Certain friends who I'm close with and have similar interests/music tastes to mine made me feel more likely to listen to a band."} [P141] \textit{``I found the recommendation for Falluah most convincing because it was liked by one of my close friends who has great taste in music."} [P51]

Disagreement, on the other hand, could lead an explanation to be less persuasive: \textit{``Sometimes I judged the artist solely based on which friend liked it. If it was a friend that I did not think I would have similarly music taste too, then I immediately ruled the artist out which may be an incorrect judgment.''} [P15]

\begin{table}
\centering
\begin{tabular}{lc} 
\toprule
\textbf{Answer Theme} & \textbf{Prevalence (\%)} \\ 
\midrule
Artist Name and Cover & 10  \\ 
Expert Friends & 12 \\ 
Popular Among Friends & 12 \\
Similar Friends & 18 \\ 
Good Friends & 26 \\ 
Overall Popularity & 13 \\
None & 9 \\ 
\bottomrule
\end{tabular}
\caption{Answer themes and their prevalence for the kinds of information participants found most convincing. Some of these were explicitly shown (e.g., overall popularity), while others were raised by participants (e.g., friends having similar taste in music, or perceived to be experts).}
\label{tab-survey2}
\end{table}

\paragraph*{Popularity only matters if people identify with the crowd}$ $
\indent People were more divided about the efficacy of popularity-based explanations.  For some, social proof was clearly an important influence: \textit{``The recommendations that had more `likes' were most powerful. I assume that there is a reason that so many people like that music.''} [P172]

This is particularly true when people see the crowd as providing useful information, as with this person who found recommendations through his friends: \textit{``The recommendations that were most convincing to me were the ones that displayed that a decent number of my friends listened to or liked the artist. I often like to hear my friends' feedback on certain artists and music tastes so that I might get a better idea of what is out there that I might like as well.''} [P32]

However, when people don't see their friends as informative for them, they dismissed friend count information: \textit{``Me and my friends' music tastes rarely match up, so I've learned to not care about what music my friends like. Since I mostly listen to mainstream music that means that I would more likely listen to artists with more likes.''} [P96]

\subsection{How important are social explanations in decision-making?}
\label{sec-diff-effects}
We have seen that different kinds of social explanations are differently persuasive, and further, that there is variation between individuals in how useful they find different kinds of social explanations.  We now look at \textbf{[RQ2]:} How do explanations interact with a person's inherent biases or preferences? 

\paragraph*{People are differently susceptible to social explanation}$ $
\indent Table~\ref{tab-survey1} shows three main groups that emerged when we asked people how they felt about the social explanations and coded their responses.  On balance, people felt that social explanations could influence their decisions about artists, but the amount of influence varied quite a bit between people.

\begin{table}
\centering
\begin{tabular}{lc} 
\toprule
\textbf{Answer Theme} & \textbf{Prevalence (\%)} \\ 
\midrule
Helped make decision & 34 \\ 
Useful information & 40 \\ 
No use or influence & 20\\ 
Other & 6\\ 
\bottomrule
\end{tabular}
\caption{Answer themes and prevalence for how much participants thought they were influenced by social explanations overall.  On balance, people saw them as presenting some useful information, though the amount of influence varied.}
\label{tab-survey1}
\end{table}

As with their reactions to particular kinds of explanation, the differences appear to hinge on whether people expect the social information to be informative: \textit{``I think that it influenced my choice on the degree to which I thought I would search the artist and how confident I felt in that decision. If I knew the person well, trusted them, or was friends with them, or if a lot of my Facebook friends liked that artist, I was definitely more likely to think about researching the artist and feeling confident about it.''} [P22]

\paragraph{Social explanation is only part of the story}
Although not cited as important as the social information, the artist's name and photo had an effect too: \textit{``What influenced me the most was the picture associated with the band or artist.''} [P66] 

For most (Table~\ref{tab-survey2}), social explanations were useful, but they were just a part of a story in which other factors also mattered: \textit{``The albums with the most interesting picture, or interesting name, with a lot of likes. If the name struck me, such as `Formidable Joy', I found myself wondering more. If a lot of my friends liked it, it must be good!''} [P7]

And, as we saw earlier with friends who had incompatible tastes, people would sometimes combine social explanation with artist information in order to reject a recommendation: \textit{``The recommendations didn't really convince me that much. It more mattered what my interests were, not my friends'.  If anything, some of the recommendations convinced me not to look up the bands; if the artist looked like a rapper, and the kid who suggested it was a younger boy from my high school who thinks he is cool I was positive that I was not going to look it up.''} [P59]

\paragraph{Explanations are a second order effect}
Our final observation is that, based on our data, explanations are a second order effect.  The standard deviations for likelihood rating shown in Table \ref{tab-likely_rat} were high and the effect size is small (Cohen's $d \approx 0.2$) even between the most and least persuasive social explanation strategies, \gfr\ and \rfr.  This suggests that other factors play an important role in people's decision-making around recommendations.

Participants' responses comments confirmed that the effect of explanations may depend on pre-conceived notions of quality, or prior information:
\textit{``Recommendations of artists that seemed established AND were endorsed by people who I respect were the most powerful. Even if they were endorsed by someone I know and respect, if they seemed to be a garage band, I did not find the recommendation powerful.''} [P117]
\textit{``I tended to find the most powerful recommendations were the ones whose genre I knew in advance and were liked by my Facebook friends that were closest to me.''} [P132]

\begin{figure}
\centering
\includegraphics[scale=0.45]{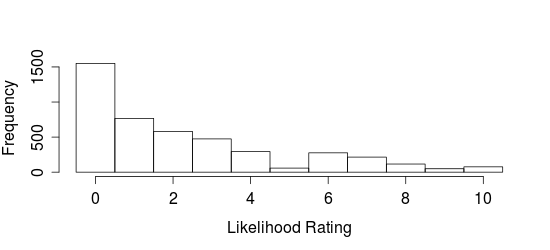}
\caption{Overall distribution of likelihood ratings across explanation strategies.  The mode is 0; frequencies decrease thereafter except for the anomalous 5 and a bump around 6.}
\label{fig-likely_distr}
\end{figure}

Further evidence is provided by the distribution of likelihood ratings (Figure~\ref{fig-likely_distr}), which shows that most ratings are below 2.  This trend is consistent across explanation strategies, which suggests that in addition to explanation, underlying every rating there is a base decision process, that on average, leans towards rejection. 

\begin{table}
\centering
\begin{tabular}{lc} 
\toprule
\textbf{Explanation} & \textbf{Fraction $>5$}\\ 
\midrule
\cfr & 0.137 \\ 
\rfr & 0.141\\ 
\nofr & 0.175\\ 
\gfr & 0.200 \\ 
\gcfr & 0.239\\
\bottomrule
\end{tabular}
\caption{Fraction of likelihood ratings above 5 (neutral rating) for each explanation strategy. Good friends-based strategies have higher fractions of ratings above 5.}
\label{tab-likely_rat_5}
\end{table}

\begin{comment}
\paragraph*{Surety}
We also asked users how sure they were. No major difference between the five explanation strategies. Users tend to more sure about the extreme ratings, as shown in Fig.~\ref{fig-sure_hist}.
\end{comment}

\begin{figure*}[t]
\centering
\includegraphics[scale=0.625]{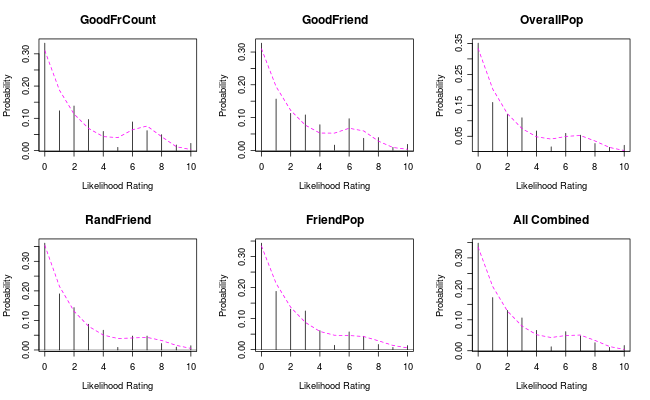}
\caption{Likelihood densities for different explanation strategies. Note how \gcfr\ and \gfr\ have higher bumps after 5 than others. The line plot shows the fit of our proposed mixture model.}
\label{fig-likely_distr_fitted}
\end{figure*}

\section{A generative model for likelihood ratings}
\label{sec-gen-model}
In this section we address \noindent \textbf{[RQ3]:} How can we model the effect of explanations on likelihood ratings?  Figure~\ref{fig-likely_distr_fitted} shows the overall distribution of likelihood ratings, along with the distribution for each social explanation strategy.  Although \gcfr\ and \gfr\ have a higher proportion of likelihood ratings over 5 (see Table~\ref{tab-likely_rat_5}), it's clear that no matter which explanation strategy is used, people have an underlying model of likelihood that has a stronger influence on their ratings than explanations.  This also came out through people's comments in Section~\ref{sec-diff-effects}.

Both the graphs and the comments suggest that a mixture model for the ratings might be appropriate, thus, we assume that a person's likelihood rating is derived from a probability distribution that is a mixture of two independent distributions.  One represents her inherent likelihood estimate for the item, and the other describes the effect of the social explanation. The density function $h$ for the ratings can be written as:
$$h(x)=af(x)+ (1-a)g(x)$$
where $f(x)$ and $g(x)$ are continous density functions representing the inherent preferences and explanations respectively. We model $x$ as a continuous variable, although it is discrete in the data ($x \in \{0,1,\text{...},10\}$). $a$ is a parameter that represents the \textit{rigidness} of the underlying likelihood model, compared to explanations; the higher $a$ is, the less effect explanations have on people's decision-making.

We first specify the base likelihood model, $f(x)$, which in this case includes both a person's base preferences and the effect of showing an artist's name and photo.  Note that we are not modeling actual preferences; rather, we are estimating whether the user is likely to try out an artist.  Our data shows a large percentage of artists with very low ratings.  This is not surprising, since we chose artists that users claimed they knew little about. Thus, we model $f(x)$ as an exponentially decaying function controlled by $\alpha$, the \textit{discernment} of an individual; discerning individuals tend to give relatively few high ratings.
\begin{equation}
f(x) = \alpha e^{-\alpha x}
\end{equation}

We now turn to modeling the effect of social explanations, $g(x)$. People described how explanations with specific friends' names had both positive and negative effects, depending on their perception of that friend's usefulness as a source of information.  Those who valued popularity-based explanations mentioned how the number of people associated with an explanation helped them decide. It seems plausible that most explanations, whether names or counts, will only be average in their persuasion, as opposed to very convincing ones on either side. Thus we model the effect of explanations by a $\mu$-centered distribution, as shown in equation \ref{equ-g-of-x}. The center of the distribution gives a sense of the \textit{receptiveness} of an individual, while the standard deviation $\sigma$ represents how different explanations might affect them differently, the person's \textit{variability}.
\begin{equation}
g(x) = \frac{1}{\sqrt{2\pi}\sigma} e^{-\frac{1}{2} \frac{(x-\mu)^2}{\sigma^2}}
\label{equ-g-of-x} 
\end{equation}

Putting things together, we get the following mixture model.
\begin{equation}
h(x) = a(\alpha e^{-\alpha x}) + (1-a)(\frac{1}{\sqrt{2\pi}\sigma} e^{-\frac{1}{2} \frac{(x-\mu)^2}{\sigma^2}}) 
\end{equation}

The mean of density $h(x)$ is given by $a/ \alpha+(1-a)\mu$. Constraining the mean to be equal to the mean of the original likelihood distribution ($c$), we have
$$\alpha = \frac{a}{c - (1-a)\mu}$$

Thus, the parameters of the model are the receptiveness ($\mu$), the variability ($\sigma$), and the rigidness ($a$) of an individual. Given an artist and an explanation, a user draws her rating from the distribution $h(x)$ as a mixture of her preference and explanation models specified by the triplet ($\mu,\sigma,a$). Over a set of the user's ratings, the prevalence of a certain rating $x$ can be approximated by $h(x)$.

\subsection{Aggregate effects of explanation strategies}
We first see how well the model explains the aggregate ratings.  For the average user represented by these ratings, we fit the model parameters for ratings from each explanation strategies separately, as well as for the combined case (Figure~\ref{fig-likely_distr_fitted}).  We evaluate the fits using residual standard error.

Table~\ref{tab-fitparams} shows the fitted parameters for the different explanation strategies. First, we observe that values for $\alpha$ are very close to one another for all strategies, giving weight to the assumption of an inherent discernment parameter for the average user that does not depend on explanation strategy. \gcfr\ exhibits the lowest value of $a$, suggesting that explanations of that type influence user ratings more. The receptiveness($\mu$) and variability($\sigma$) scores together explain how \gcfr\ and \gfr\ have more ratings above 5, and hence are more consistently persuasive than the others (and giving further support to our earlier findings).

\begin{table}
\centering
\begin{tabular}{llcccc} 
\toprule
\textbf{Explanation} & \textbf{Error} & \textbf{$\alpha$}\textit{(computed)} & \textbf{$\mu$} & \textbf{$\sigma$} & \textbf{$a$} \\ 
\midrule
\cfr & 0.022 & 0.44 & 6.85 & 3.61 & 0.74 \\ 
\rfr & 0.018 &0.49 & 7.10 & 3.57 & 0.71\\ 
\nofr & 0.026 & 0.49 & 6.89 & 3.10 & 0.66\\ 
\gfr & 0.030 & 0.46 & 6.46 &2.51 & 0.66\\ 
\gcfr & 0.034 & 0.50  & 6.84 & 2.26 & 0.61\\ 
Combined &0.022 & 0.47 & 6.88 & 3.05 & 0.69\\ 
\bottomrule
\end{tabular}
\caption{Fit parameters for likelihood densities of different explanation strategies. \gcfr\ has the lowest rigidness ($a$), which suggests people were more swayed by this explanation strategy.}
\label{tab-fitparams}
\end{table}

\begin{comment}
\begin{figure}[t]
\label{fig-likely_distr}
\centering
\includegraphics[scale=0.45]{graphs/hist_likely2}
\caption{Likely rating distribution}
\end{figure}
\end{comment}

\subsection{Different users, different models}
\label{subsec_clusters}
\begin{figure}[t]
\centering
\includegraphics[scale=0.4]{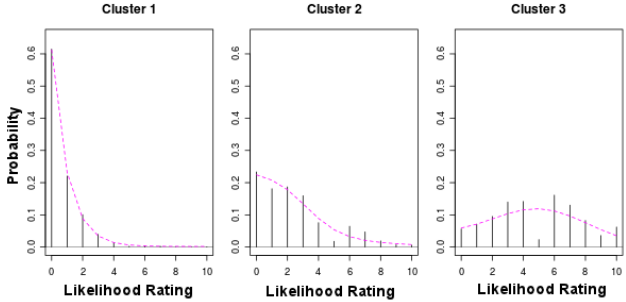}
\caption{Likelihood rating distributions for three clusters of users. These distributions bring out the three types of users: ones on whom explanations had no effect, those who found them useful and those who relied on them more heavily. As before, the line plots show the fitted mixture models.}
\label{fig-likely_distr_cluster}
\end{figure}

Until now, we have analyzed the distribution of the aggregate population.  However, as we saw earlier, people are differently affected by explanations; we now look at how we might refine the models by exploiting the differences in susceptibility to explanations demonstrated by Table~\ref{tab-survey1}.
To do this, we group users into three clusters using a standard k-means algorithm, representing users by their mean and variance of ratings. The mean ratings in the three computed clusters are 0.67, 2.44, and 4.89 respectively. Figure~\ref{fig-likely_distr_cluster} shows the distribution of likelihood ratings for the three clusters, and Table~\ref{tab-fitparamscluster} shows the fitted parameters (we do not fit for individuals for fear of overfitting, since users have about 30 ratings).   

The plots give evidence of these three types of users in the data, with cluster 1 roughly representing the ``no use or influence'' case, cluster 2 representing ``useful information'', and cluster 3 representing ``helped make decision''. Parameter $a$ decreases from cluster 1 to 3, suggesting the decreasing rigidness of individuals towards explanations. Clusters 1 and 3 serve as composing examples of the mixture model: cluster 1 illustrates the dominance of the exponential distribution, while cluster 3 is highly gaussian.   

\begin{table}
\centering
\begin{tabular}{ccclcrc} 
\toprule
\textbf{Cl\#} & \textbf{N} & \textbf{Ratings} & \textbf{Error} & \textbf{$\mu$} & \textbf{$\sigma$} & \textbf{$a$} \\ 
\midrule
1 & 89 & 1817 & 0.001 & 0.05 & 78.82 & 0.62\\ 
2 & 84 & 1610 & 0.01  & 1.43 & 1.98 & 0.50\\ 
3 & 64 & 1119 & 0.04  & 4.99 & 3.22 & 0.08\\ 
\bottomrule
\end{tabular}
\caption{Fitted parameters for three clusters of users. The effect of explanations increases from Cluster 1 to 3, as shown by the values for $a$.}
\label{tab-fitparamscluster}
\end{table}

\paragraph{Personalization} In Section~\ref{sec-diff-effects}, we observed how people are differently susceptible to social explanation. The above data provides weight to that observation, and opens up opportunities for personalization of explanations.  In a practical system, this could be done in multiple stages.  When users first join the system, they can be assigned population averages for these parameters for each explanation strategy.  As they encounter explanations, their preferences can be either explicitly captured (e.g., through rating whether an explanation is helpful, as with Amazon reviews) or implicitly inferred based on their reaction to the explained recommendation.  As we build up data, we can compare them to cluster models such as those described here to see whether explanations are helpful at all, or have individual models for each user.  Eventually, we can infer which types of explanations are the most appropriate for an individual user and prefer showing them when possible.

\begin{figure}[t]
\centering
\includegraphics[scale=0.30]{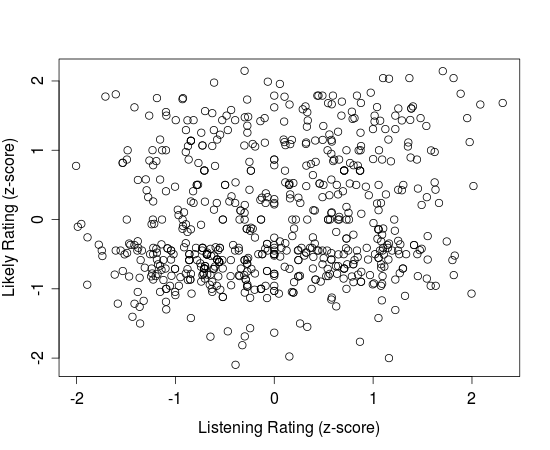}
\caption{Z-scores of likelihood and listening ratings. The two ratings show little correlation (correlation coeff=0.17)}
\label{fig-likely_listen}
\end{figure}

\begin{table}[t]
\centering
\begin{tabular}{lccccc} 
\toprule
\textbf{Explanation} & \textbf{N} & \textbf{Mean} & \textbf{Std. Dev.} & \textbf{Mean-likely}\\ 
\midrule
\cfr & 190 & 4.14 & 2.85 & 2.12\\ 
\rfr & 192 & 4.57 & 3.09 & 2.08 \\
\nofr & 198 & 4.86 & 2.92 & 2.36\\ 
\gfr & 133 & 4.57 & 2.86 & 2.52\\ 
\gcfr & 122 & 4.63 & 2.84 & 2.71 \\ 
\bottomrule
\end{tabular}
\caption{Listening ratings for artists, binned by explanation strategy. \nofr\ performs the best in Phase II, but we found no significant difference between the ratings.}
\label{tab-listen_rat}
\end{table}

\section{Phase II: Consumption}
Having analyzed likelihood ratings, we now focus on \textbf{[RQ4]:} How effective are these explanations in directing people to items that receive high consumption ratings?  First, we study how the different explanation strategies shown in Phase I affected consumption ratings in Phase II.  We then contrast the overall consumption ratings with likelihood ratings.

\subsection{Do explanations affect consumption ratings?}
Table~\ref{tab-listen_rat} shows the consumption ratings for different explanation strategies.  We note that the means for consumption are higher than for likelihood. While \gcfr\ performed best for likelihood, we find that \nofr\ records the highest mean for consumption. However, we must be careful with making conclusions since (except for \cfr), the means for different strategies are quite close, and an ANOVA with repeated measures confirms the differences are not significant ($F(4,378)=1.64, p=0.2$).

\begin{figure}[t]
\centering
\includegraphics[scale=0.25]{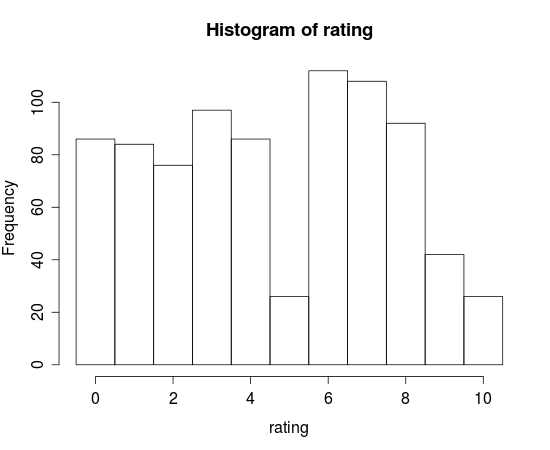}
\caption{Distribution of consumption ratings for all users. Apart from very high ratings \{9,10\} and the anomalous 5, ratings are evenly distributed.}
\label{fig-rating_distr}
\end{figure}

Since \nofr, \gcfr\, and \gfr\ all have comparable ratings, this implies that given a delay of a few days, explanations lose their influence on a user's decision. Figure~\ref{fig-rating_distr} shows how ratings are close to uniformly distributed across the 11-point scale (except very high ratings, $>8$ which show a dip, and the anomalous 5). The different explanation strategies exhibit similar distributions. 
%Without any external effects, this is an expected result of recommendations drawn randomly from all artists. Since all artists were Liked by atleast one friend, it also confirms that overall friends' preferences can be diverse. 

\subsection{Does likelihood predict consumption?}
We next look at whether likelihood ratings can predict later consumption ratings.  Figure~\ref{fig-likely_listen} shows how the two compare, z-score adjusted to control for individual biases in numerical ratings. It is apparent that there is little correlation between likelihood and consumption ratings ($r=0.17$), suggesting that the persuasiveness and informativeness of an explanation are quite different \cite{bilgic05}.  Later we will discuss ways to increase the informativeness of explanations through presenting other information, and in the limiting case where we provide almost all the information about an item in a recommendation (such as recommending pictures), these ratings should be close together.  But our results show that these two ratings can be quite far apart, suggesting that it will be useful to think about the two kinds of rating independently.

\subsection{Modeling Likelihood and Consumption} 
\label{sec-combined-rec-framework}
As noted earlier, scenarios of two-phase recommendation are common on the web---for example, clicking a movie recommendation on Netflix and rating it after watching, or clicking a Page recommendation on Facebook and deciding to Like it.
In general, current approaches to information filtering assume that the two ratings are correlated (or have access to only one), and hence optimize only one of the rating objectives.  For example, recommender systems research focuses mainly on consumption ratings, while ad systems typically optimize click-through rates.

One way we could make use of modeling both likelihood and consumption is by conceptualizing the decision-making as a sequential process.  A user proceeds to consume an artist recommendation only after he evaluates a high enough likelihood for liking that artist. Thus we could set up an optimization framework:
$$ \text{maximize }R \textit{ s.t. } L > \epsilon_u $$
where $L$ and $R$ are the likelihood and consumption ratings for an artist respectively\footnote{Our formulation is different from multiple objective optimization \cite{rodriguez12,adomavicius11}, since the two objectives are sequential.}. $\epsilon$ can be initialized to a reasonable global value (such as 5 in our case), or a user-specific $\epsilon_u$.  Models could iteratively decrement $\epsilon$ in case enough recommendations cannot be retrieved, or depending on recommendation goals, may use alternate values for $\epsilon$.  For serendipity, one may prefer may prefer to set $\epsilon$ lower, for instance. Note that in a domain where $R$ and $L$ are highly correlated, equation reduces to the standard one-phase optimization, maximizing $R$. 

$L$ may depend on the explanation shown, in which case there will be multiple likelihood values for a single item. The models for $L$ and $R$ can be based on standard collaborative filtering models \cite{susurvey09} or socially enhanced variants \cite{ma11}.

\section{Discussion}
We find that social explanations, especially ones involving close friends, are persuasive, though they have secondary effects compared to other sources of informtion about recommended items. However, our data also shows that persuasive explanations may not be informative---that people's ratings of expected liking aren't good proxies of their actual liking of the artists.  In this section, we discuss the opportunities and questions that our findings point to, along with the limitations of our study.

\paragraph{Improving expectations of informativeness}
One major finding is that the effect of social explanations is based heavily on a user's expectations of how informative the explanation will be: how they perceive a friend's music tastes to be similar to theirs, or how much they expect to agree with the crowd.   Our explanation interfaces were fairly minimal because, as shown in Section~\ref{sec-social-wild}, many real social explanation settings---particularly those that present a list of recommended items---convey little additional information beyond a title and a social explanation.

Our results suggest that this might be a mistake, and that systems should design explanation interfaces to increase the informativeness of the explanation.  For instance, the interface could show  information about similarity to people used in social explanations, either by translating similarity metrics into legible indicators (as with some of the explanation interfaces shown in \cite{herlocker00}) or by using representative examples of items liked.  It could also show information designed to convey expertise, such as the quantity, diversity, or rarity of items an explainer likes.  Based on our results, an effective display of this kind of information might make both individual-based and crowd-based social explanations more useful.
%(or, depending on the system owner's goals, more persuasive).

\paragraph{Balancing persuasiveness and informativeness}
Our results also call the difference between persuasiveness and informativeness into sharp focus \cite{bilgic05}, showing that social explanations along with basic artist information have a limited ability to help people predict their actual liking of a recommended item.  Section~\ref{sec-combined-rec-framework} talks about one way to deal with this difference, by modeling persuasiveness and informativeness separately.  This approach corresponds to the click-through/purchase distinction in customer behavior in e-commerce sites, and it does have some advantages.  

Considering them separately gives designers more freedom to optimize users' experiences and support different recommendation goals \cite{mcnee03}. 
Our initial proposed model suggests that increasing persuasiveness might increase overall user activity and consumption, though at some risk of eroding trust if the system persuades users to consume items they don't actually like.  Systems might also effectively support serendipity by increasing the persuasiveness of explanations for items where the consumption model predicts high ratings and the likelihood model predicts low ratings.  Tuning the likelihood threshold might also support users who prefer either riskier or more conservative recommendations.  

\paragraph{Increasing informativeness of explanations}
An alternative approach to managing the gap between likelihood and consumption ratings would be to enrich explanations in order to close the gap.  Our suggestions above about increasing the informativeness of social explanation are one such strategy.  However, as we've seen, social explanations are just one part of people's decision-making process.  A number of other interface elements have been proposed that might help explain recommendations, including tags associated with the item \cite{vig09}, indicators of the systems's confidence in the recommendation \cite{mcnee03}, and the predicted rating itself \cite{cosley03}.

These interface elements fall into four main classes: tokens of the item itself (such as genres or music clips for music, or trailers, genres, and actors for a movie); data that people attach to the item (ratings, tags, reviews); metadata about those people (similarity information, their ratings); and information about the recommendation system's algorithms (confidence, predicted ratings).  Our hypothesis is that item information is more informative, and social and algorithm information are more persuasive, but this is an open question.  The space for designing explanations is rich, and more work is needed to explore the effect of these sources of information on both the persuasiveness and the informativeness of explanations of these various types.  

\paragraph{Modeling and Personalization}

Our results point towards the merit of personalizing explanation \textit{strategies}, in addition to showing personalized explanations (Section~\ref{subsec_clusters}). Our model may be extended for other explanations, since we make no assumption about the explanations except that they have a gaussian distribution of effects. For instance, in Section~\ref{sec-gen-model}, we find that some users (cluster 1, Figure \ref{fig-likely_distr_cluster}) do not seem to be affected by social explanations at all, but it is possible they may find other explanations (such as tags, genres) useful. As long as the effects are nearly gaussian, we may use the same model for those explanations too.  

There could also be variation within strategies. For example, the relative count of friends who Like an artist, or number of explicit names shown, may impact how a user perceives the explanation (though in our experiments, we did not find a correlation between friend or overall counts and likelihood). Modeling these fine-grained effects can be interesting future work.

Finally, our analysis adds further weight to the importance of \textit{interface elements} such as explanations on how users evaluate a recommendation. Through our model and recommendation framework, we take the first steps towards modeling these effects. In general, it can be useful to augment recommender systems by including these additional signals, either as new item features or through novel models.

\begin{comment}
\begin{table}
\centering
\begin{tabular}{|c|l|l|} \hline
Answer Category & Prevalence & Example \\ \hline
Comfortable & 78 & Place \\ \hline
Not Comfortable & 19 & Place\\ \hline
Others & 3 & Place\\ 
\hline
\end{tabular}
\caption{How comfortable are you seeing your friend name (and yours by extension)?}
\label{tab-survey3}
\end{table}
\end{comment}

\paragraph{Acceptability of social explanation}
Social explanations involve disclosing personal information to friends or even strangers. While we discussed them mainly from the perspective of utility, we have to be mindful of any social norms or privacy expectations these explanations might violate, or personal information they might disclose \cite{kosinski13}.

At least for the music domain, most participants we surveyed seemed to be comfortable with the idea of sharing Likes: \textit{``Yes, it was just interesting to see which of my friends liked which artists. Depending on how well I knew the person, or what kind of music they listened to, I was more open to listening to the artist.''} [P4]

This may be because people are already used to having their information public in online settings: \textit{``While it is offputting that so much information is online, I do know that this information is accessible whether you show me or not, so I got over it pretty quickly.''} [P82]

However, some were still alarmed by the information that can be disclosed. \textit{``No, I was not totally comfortable.  Since it could take my friends' information, it could take mine and share it.  It felt like a breach of privacy.''} [P100]

In particular, social explanations have the potential to misrepresent a person's preferences, taking them out of context by associating him with one or two of his Likes:
\textit{``I suppose---I'm not sure that they would be pleased to know that their information appeared in this kind of way. Music preferences are very personal, and this kind of exercise may tend to pull one `Liked' artist out of context the user's general preferences.''} [P128]

Overall, however, participants did not view privacy as a major issue and saw social explanation as a generally useful and acceptable thing to do, at least in this domain.  Still, this may not be true for the overall population and for other domains, so exploring design choices that allow users to disable explanation strategies or restrict usage of their data for explanation may be a good idea. This would have little effect on the system's ability to make good explanations in practice but have benefits for people's perceptions of having control over the system.

\paragraph{Limitations}
We want to point out five main factors around our study that readers should bear in mind when applying our results.  First, our users are fairly young and primarily drawn from a single university.  Older users might have different perceptions of the usefulness and acceptability of social explanations.  Second, we focused on the music domain.  This was intentional, to support the collection of consumption ratings, but does mean that our results may not apply in domains where consuming items is more costly in terms of money or time. Third, although we took care not to include artists familiar to a user, they were all chosen from her friends' Likes. This might have introduced a selection bias, especially if a few friends Liked most of the artists. 
Fourth, although we chose a representative sample of social explanation strategies, we did not cover the entire space.  Interfaces might show multiple names, or combine other sources of social information.  
Finally, we focused on explanations in recommendation list interfaces that show relatively little information. Exploring how people make sense of an Amazon or Best Buy product page (with richer information including detailed item information and explanations such as rating histograms) is a largely open, but interesting, question.

\section{Conclusion}
Still, our results add to knowledge around the effects of social explanations on user preferences, both before and after consumption of a recommendation.  Based on our findings, we presented a generative model that explains much of the variation in likelihood ratings and that can be personalized. The low correlation between likelihood and consumption ratings highlights another facet of how users make sense of explanations, that persuasiveness and informativeness of an explanation are largely independent. This suggests that modeling one may not be sufficient, and we proposed an optimization framework that can be useful for thinking about both likelihood and consumption ratings.

Going forward, we believe that explanations, and other external informational elements, influence the evaluation of recommendations in non-trivial ways.  They raise interesting questions at the intersection of user interfaces and recommender systems. For example, which types of explanations are both persuasive and informative, and for which users? Explicitly modeling interface elements could provide a basis for design choices at the interface level, as well as help in improving the perceived quality of a recommender system.

\section{Acknowledgments}
This work was supported by the National Science Foundation under grants IIS 0910664 and IIS 0845351. Michael Triche helped in developing the experimental system.

\bibliographystyle{abbrv}

\end{document}